\documentclass[twocolumn,showpacs,preprintnumbers,amsmath,amssymb]{revtex4}

\usepackage{graphicx}
\usepackage{bm}

\begin{document}


\title{The spin-dependent semiconductor Bloch equations: a microscopic theory of Bir-Aronov-Pikus spin-relaxation}

\author{C. Lechner}
\author{U. R\"ossler}
\affiliation{Institut f\"ur Theoretische Physik, Universit\"at Regensburg,
             D-93040 Regensburg}
\date{\today}
\begin{abstract}
Semiconductor Bloch equations, in their extension including the spin degree of freedom of the carriers, are capable to describe spin dynamics on a microscopic level. In the presence of free holes, electron spins can flip simultaneously with hole spins due to electron-hole exchange interaction. This mechanism named after Bir, Aronov and Pikus, is described here by using the extended semiconductor Bloch equations \cite{Roessler:2002} and considering carrier-carrier interaction beyond the Hartree-Fock truncation. As a result we derive microscopic expressions for spin-relaxation and spin-dephasing rates. 
\end{abstract}


\maketitle


Semiconductor Bloch equations (SBE) are a well established concept to describe the dynamics of carriers in a semiconductor or quantum structure by a scalar light field.\cite{Lindberg:1988,Haug:1993,Khitrova:1999} It has been used successfully in modelling the time evolution due to carrier-carrier interaction on different time scales including the \emph{coherent} and the \emph{relaxation} regime.\cite{Binder:1995} Thus, SBE have become the dominating tool in the theory of semiconductor lasers and in designing even the complex structures of \emph{vertical cavity surface emitting lasers} (VCSEL). One phenomenon, however, connected with VCSELs points to a deficiency of the SBE: these laser structures are known for their polarization instability, i.e., the uncontrolled switching of the laser modes between the two possible transverse polarizations.\cite{Choquette:1993,SanMiguel:1995,Ando:1998} In addition, the investigation of semiconductor quantum structures as model systems for coupled Rabi oscillations with electrons, heavy- and light-hole states (each spin-degenerate) required to extend the two-level SBE to six-level SBE and to take into account the polarization degree of freedom of the exciting light.\cite{Binder:2000} 

More recently, the carrier spin and its dynamics have gained much interest in the field of spintronics.\cite{Zutic:2004} Spin dynamics in semiconductors \cite{Meier:1984} and quantum structures,\cite{Sham:1993,Awschalombook:2002} formulated so far in a more phenomenological way, is ruled by different mechanisms: one of which is related to the electron-hole exchange interaction.\cite{Denisov:1973} It becomes relevant if the semiconductor system contains besides electrons also holes (e.g., due to doping or high excitation). This \emph{Bir-Aronov-Pikus} (BAP) mechanism, originally considered for bulk semiconductors,\cite{Bir:1975} has been described also for semiconductor quantum structures,\cite{Sham:1993,Maialle:1997} but never by a rigorous microscopic treatment of the spin dynamics. In this perspective, the SBE have been formulated for the six-level system,\cite{Roessler:2002} considering spin-splitting of the electronic energies due to spin-orbit coupling caused by bulk inversion (BIA) \cite{Dresselhaus:1955} or structure inversion asymmetry (SIA).\cite{Rashba:1960} These extended SBE were designed only within the \emph{Hartree-Fock truncation} leading to the coherent regime, thus neglecting scattering processes, responsible for relaxation. Recently, we have used the extended optical Bloch equations (SBE without carrier-carrier interaction \cite{Schaefer:2002}) to provide a microscopic approach to the longitudinal ($T_1$) and transverse ($T_2$) relaxation times due to electron-phonon interaction.\cite{Lechner:2005} In this approach we have considered scattering between electrons and phonons in second order Born approximation to provide a microscopic formulation for the D'yakonov Perel' (DP) mechanism of spin-relaxation.\cite{Dyakonov:1972} The analogous concept is applied here to the electron-hole exchange interaction and yields the microscopic formulation of spin-relaxation due to the BAP mechanism.


In the following, we concentrate on spin dynamics in a semiconductor quantum well (QW) under excitation with circularly polarized light leading to a nonequilibrium spin distribution due to optical selection rules. Let the system be described by the Hamiltonian 
\begin{equation}
\mathcal{H} = \mathcal{H}_{0} + \mathcal{H}_{\text{light}} + \mathcal{H}_{\text{coul}}\;,
\label{sysham}
\end{equation}
where $\mathcal{H}_{0}$ is the kinetic part including BIA and SIA spin-orbit coupling, $\mathcal{H}_{\text{light}}$ the interaction with the exciting light field, and $\mathcal{H}_{\text{coul}}$ the Coulomb interaction between the carriers. We adopt the notation of our previous works \cite{Roessler:2002,Lechner:2005} and use the basis in which the kinetic part of the Hamiltonian for the six-level system is \emph{diagonal}
\begin{eqnarray}
\nonumber
\mathcal{H}_{0} &=& \sum_{\mathbf{k}'\, m_c'} \epsilon_{m_c'}(\mathbf{k}')\, c^{\dagger}_{m_c'}(\mathbf{k}')\, c_{m_c'}(\mathbf{k}')\\
&& + \sum_{\mathbf{k}'\, m_v'} \epsilon_{m_v'}(\mathbf{k}')\, v_{m_v'}(\mathbf{k}')\, v^{\dagger}_{m_v'}(\mathbf{k}')\;.
\end{eqnarray}
Here, $c_{m_c}(\mathbf{k})$ $\big[v_{m_v}(\mathbf{k})\big]$ are fermion operators for electrons (light- and heavy-holes) with spin quantum numbers $m_c=\pm1/2$ ($m_v=\pm 1/2\, , \pm3/2$) defined with respect to the in-plane wave vector $\mathbf{k}$. The time dependence of the operators is understood. The single particle energies $\epsilon_{m_c'}(\mathbf{k}')$ $\big[\epsilon_{m_v'}(\mathbf{k}')\big]$ describe subbands, which are spin-split due to spin-orbit interaction. In dipole approximation, the interaction with the light field reads 
\begin{eqnarray}
\nonumber
\mathcal{H}_{\text{light}} &=& - \sum_{\substack{m_c'\, m_v' \\ \mathbf{k}'}} \Big[\mathbf{E}(t)\cdot \mathbf{d}_{m_c'\, m_v'}(\mathbf{k}')\, c^{\dagger}_{m_c'}(\mathbf{k}')\, v^{\dagger}_{m_v'}(\mathbf{k}')
\\
&&+ \mathbf{E}^{*}(t)\cdot \mathbf{d}^{*}_{m_c'\, m_v'}(\mathbf{k}')\, v_{m_v'}(\mathbf{k}')\, c_{m_c'}(\mathbf{k}') \Big]\; ,
\end{eqnarray}
where $\mathbf{E}(t)$ is the electric field vector and $\mathbf{d}_{m_c'\, m_v'}(\mathbf{k}')$ is the dipole matrix element connecting valence and conduction band states (for details see Ref.~\onlinecite{Roessler:2002}). 

The carrier-carrier interaction can be split up into four parts 
\begin{eqnarray}
\mathcal{H}_{\text{coul}}=\mathcal{H}_{\text{ee}} + \mathcal{H}_{\text{hh}} + \mathcal{H}^{\text{C}}_{\text{eh}} + \mathcal{H}^{\text{X}}_{\text{eh}}\; .
\label{Hamcoul}
\end{eqnarray}
Here, $\mathcal{H}_{\text{ee}}$ ($\mathcal{H}_{\text{hh}}$) describes the Coulomb interaction between electrons (holes) in the conduction (valence) band. The remaining terms account for electron-hole interaction, the \emph{direct} Coulomb term $\mathcal{H}^{\text{C}}_{\text{eh}}$ and the \emph{exchange} term $\mathcal{H}^{\text{X}}_{\text{eh}}\,$.\cite{Denisov:1973} In the frame of SBE,\cite{Haug:1993,Khitrova:1999,Binder:1995} especially for the coherent regime, carrier-carrier interaction has been considered so far only with respect to renormalization of the single-particle energies and of the interaction with the light field, while the electron-hole exchange has been ignored. However, it is just $\mathcal{H}^{\text{X}}_{\text{eh}}$ which can cause spin-flips, thus contributing to the spin-dynamics due to the BAP mechanism and, hence, is of interest here. As derived in Ref. \onlinecite{Roessler:2002}, the exchange term reads
\begin{widetext}
\mbox{}\\
\vspace{-0.6cm}
\begin{eqnarray}
\mathcal{H}^{\text{X}}_{\text{eh}} &=& 
\frac{1}{2}\, \sum_{\substack{m_{c}\,m_{c}'\\ m_{v}\, m_{v}'}}\, \sum_{\substack{\mathbf{k}\, \mathbf{k}' \\ \mathbf{q}}} \mathcal{V}^{\text{X}}_{m_{c}\, m_{v}\, m_{c}'\, m_{v}'}(\mathbf{k},\, \mathbf{k}',\, \mathbf{q})\, c^{\dagger}_{m_{c}}(-\mathbf{k}+\mathbf{q})\, c_{m_{c}'}(-\mathbf{k}'+\mathbf{q})\, v^{\dagger}_{m_{v}'}(\mathbf{k})\, v_{m_{v}}(\mathbf{k}') \;.
\end{eqnarray}
\end{widetext}
The detailed form of the interaction matrix element $\mathcal{V}^{X}_{m_{c}\, m_{v}\, m_{c}'\, m_{v}'}(\mathbf{k},\, \mathbf{k}',\, \mathbf{q})$ will not become important in the following. But we emphasize, that the structure of this matrix element makes simultaneous flips of electron and hole spins possible, which, in the electron system, finally contribute to spin-relaxation.\cite{Roessler:2002}

While the dynamics of the whole system is contained in the equations of motion (EOM) of the $6 \times 6$ density matrix, we concentrate here on the dynamics of the electron spins by looking at the EOM of the $2 \times 2$ density matrix for the electron subsystem 
\begin{equation}
\bm{\varrho}^{(m_c\, \bar{m}_c)}(\mathbf{k}) = 
\left(\begin{array}{c c}
\varrho_{m_c\, m_c}(\mathbf{k}) & \varrho_{m_c\, -m_c}(\mathbf{k}) \\
\varrho_{-m_c\, m_c}(\mathbf{k}) & \varrho_{-m_c\, -m_c}(\mathbf{k}) 
\end{array} \right)\;.
\end{equation}
The single entries are expectation values of products of a creation and an annihilation operator $\varrho_{m_c\, \bar{m}_c}(\mathbf{k})= \langle c^{\dagger}_{m_c}(\mathbf{k})\, c_{\bar{m}_c}(\mathbf{k}) \rangle\,$. Their EOM read
\begin{widetext}
\mbox{}\\
\vspace{-0.6cm}
\begin{eqnarray}
\nonumber
i\, \hbar \partial_t {\varrho}_{m_c\, \bar{m}_c}(\mathbf{k}) &=& \Big[\epsilon_{m_c}(\mathbf{k}) - \epsilon_{\bar{m}_c}(\mathbf{k})\Big]\,\varrho_{m_c\, \bar{m}_c}(\mathbf{k}) + \sum_{m_v}\Big[\mathbf{E}(t)\cdot \mathbf{d}^{cv}_{\bar{m}_c\, m_v}\, P_{m_c\, m_v}(\mathbf{k}) - \mathbf{E}^*(t)\cdot \mathbf{d}^{cv\,*}_{m_c\, m_v}\, P^{\dagger}_{\bar{m}_c\, m_v}(\mathbf{k})\Big]\\ \nonumber
&&- \sum_{\bar{\mathbf{k}}\, \mathbf{q}} \sum_{\substack{m_{c}' \\ \tilde{m}_{v} \, \tilde{m}_{v}'}}
\Big[\mathcal{V}^{\text{X}}_{\bar{m}_{c}\, \tilde{m}_v\, m_c'\, \tilde{m}_{v}'}(-\mathbf{k} + \mathbf{q},\, \bar{\mathbf{k}},\, \mathbf{q})\,  \langle c^{\dagger}_{m_{c}}(\mathbf{k})\, c_{m_c'}(-\bar{\mathbf{k}} + \mathbf{q})\, v^{\dagger}_{\tilde{m}_v}(-\mathbf{k}+ \mathbf{q})\, v_{\tilde{m}_{v}'}(\bar{\mathbf{k}})\rangle\\
&& \mbox{} \hspace{1.8cm}- \mathcal{V}^{\text{X}}_{m_{c}'\, \tilde{m}_v\, m_c\, \tilde{m}_{v}'}(\mathbf{\bar{k},\, -\mathbf{k}+\mathbf{q},\, \mathbf{q}}) \,\langle c^{\dagger}_{m_{c}'}(-\bar{\mathbf{k}}+ \mathbf{q})\, c_{\bar{m}_c}(\mathbf{k})\, v^{\dagger}_{\tilde{m}_v'}(\bar{\mathbf{k}})\, v_{\tilde{m}_{v}}(-\mathbf{k} + \mathbf{q})\rangle\Big] \; ,
\label{eomoffuntrunc}
\end{eqnarray}
\end{widetext} 
where we have introduced the interband polarization $P_{m_c\, m_v}(\mathbf{k})=\langle c^{\dagger}_{m_c}(\mathbf{k})\,v^{\dagger}_{m_v}(\mathbf{k})\rangle$.\cite{Haug:1993} Due to the many-body contributions, the dynamics of $\varrho_{m_c\, \bar{m}_c}(\mathbf{k})$ are ruled by four-point density matrices and, consequently, we run into a \emph{hierarchy problem}, which can be solved by an appropriate truncation. The \emph{Hartree-Fock} (HF) truncation scheme \cite{Haug:1993}, as used in Ref. \onlinecite{Roessler:2002}, factorizes the expectation values of the four-operator terms into a product of two-operator terms under the condition that they are macroscopic, namely, either electron (hole) densities or polarizations. While closing the hierarchy and renormalizing the eigenenergies and the dipole interaction the HF truncation limits the EOM to the \emph{coherent} regime, because no scattering processes are taken into account. In order to include these processes, which are essential for spin-relaxation and -dephasing, we go beyond the HF truncation by considering the \emph{reduced four-operator terms} \cite{Haug:1993}, defined as the difference between the expectation value of the untruncated four-operator term and its HF truncated product. For $\langle c^{\dagger}_{m_{c}'}(-\bar{\mathbf{k}}+ \mathbf{q})\, c_{\bar{m}_c}(\mathbf{k})\, v^{\dagger}_{\tilde{m}_v'}(\bar{\mathbf{k}})\, v_{\tilde{m}_{v}}(-\mathbf{k} + \mathbf{q})\rangle$ [see Eq. \eqref{eomoffuntrunc}] it reads
\begin{widetext}
\mbox{}\\
\vspace{-0.6cm}
\begin{eqnarray}
\nonumber
\delta\,\langle c^{\dagger}_{m_{c}'}(-\bar{\mathbf{k}}+ \mathbf{q})\, c_{\bar{m}_c}(\mathbf{k})\, v^{\dagger}_{\tilde{m}_v'}(\bar{\mathbf{k}})\, v_{\tilde{m}_{v}}(-\mathbf{k} + \mathbf{q})\rangle &=&
\langle c^{\dagger}_{m_{c}'}(-\bar{\mathbf{k}}+ \mathbf{q})\, c_{\bar{m}_c}(\mathbf{k})\, v^{\dagger}_{\tilde{m}_v'}(\bar{\mathbf{k}})\, v_{\tilde{m}_v}(-\mathbf{k} + \mathbf{q})\rangle \\
&&- \langle c^{\dagger}_{m_{c}'}(\mathbf{k})\, c_{\bar{m}_c}(\mathbf{k}) \rangle \langle  v^{\dagger}_{\tilde{m}_v'}(-\mathbf{k}+ \mathbf{q})\, v_{\tilde{m}_v}(-\mathbf{k} + \mathbf{q})\rangle\, \delta_{\mathbf{k},\, -\bar{\mathbf{k}} + \mathbf{q}}\;.
\label{hfcon} 
\end{eqnarray}
\end{widetext}

The scattering contributions are found by solving the EOM of the reduced four-operator terms which contain the \emph{complete} information about the scattering in expectation values of four- and six-operator terms. In analogy to the case of electron-phonon scattering \cite{Lechner:2005} we truncate these terms by factorizing them into their macroscopic parts and taking into account only those, which contribute in \emph{second order Born approximation}. After integrating the arising equations and applying the \emph{adiabatic} and the \emph{Markov} approximation \cite{Kuhn:1992}, we achieve a closed set of equations for the reduced four-operator terms, which can be solved and used in Eq. \eqref{eomoffuntrunc} (for technical details see Ref. \onlinecite{Lechner:2005}). Thus, the EOM for the diagonal entries of the $2 \times 2$ density matrix due to the electron-hole exchange scattering can be cast into the form
\begin{eqnarray}
\nonumber
\partial_t{\varrho}_{m_c\,m_c}(\mathbf{k})\pmb{\vert}_{\text{X}} 
&=& \Gamma^{\text{out}\, \text{X}}_{m_c\,m_c}(\mathbf{k})\,\varrho_{m_c\,m_c}(\mathbf{k})\\ && +\Gamma^{\text{in}\, \text{X}}_{m_c\,m_c}(\mathbf{k})[1-\varrho_{m_c\,m_c}(\mathbf{k})]\;,
\label{T11}
\end{eqnarray}
with $\Gamma^{\text{out}\, \text{X}}_{m_c\,m_c}(\mathbf{k})$ $\left[\Gamma^{\text{in}\, \text{X}}_{m_c\,m_c}(\mathbf{k})\right]$ accounting for the exchange scattering out of (into) the state with spin $m_c$ at wave vector $\mathbf{k}$. The derivation of the scattering contributions for the different four-operator terms in Eq. \eqref{eomoffuntrunc} follows the same scheme. Thus, we present here only the results for the reduced four-operator term $\langle c^{\dagger}_{m_{c}'}(-\bar{\mathbf{k}}+ \mathbf{q})\, c_{m_c}(\mathbf{k})\, v^{\dagger}_{\tilde{m}_v'}(\bar{\mathbf{k}})\, v_{\tilde{m}_{v}}(-\mathbf{k} + \mathbf{q})\rangle$. The corresponding out-scattering rate $\Gamma^{\text{out}\, \text{X}}_{m_c\,m_c}(\mathbf{k})$ reads
\begin{widetext}
\mbox{}\\
\vspace{-0.6cm}
\begin{eqnarray}
\nonumber
\Gamma^{\text{out}\, \text{X}}_{m_c\,m_c}(\mathbf{k}) &=& 
\frac{2\, \pi}{\hbar} \sum_{\bar{\mathbf{k}}\, \mathbf{q}}\sum_{\substack{m_c'\\  \tilde{m}_v\,\tilde{m}_v'}} |\mathcal{V}^{\text{X}}_{m_c\,\tilde{m}_v'\, m_c'\, \tilde{m}_v}(-\mathbf{k}+\mathbf{q},\, \bar{\mathbf{k}},\, \mathbf{q})|^2\, \delta[\epsilon_{m_c}(\mathbf{k}) - \epsilon_{\tilde{m}_v}(\mathbf{k}- \mathbf{q}) - \epsilon_{m_c'}(-\bar{\mathbf{k}} + \mathbf{q}) + \epsilon_{\tilde{m}_v'}(-\bar{\mathbf{k}})]  \\
&&\mbox{} \hspace{1.8cm}\times [1 - \varrho_{\tilde{m}_v'\,\tilde{m}_v'}(-\bar{\mathbf{k}})]\, [1 - \varrho_{m_c'\,m_c'}(-\bar{\mathbf{k}} + \mathbf{q})]\, \varrho_{\tilde{m}_v\, \tilde{m}_v}(\mathbf{k}- \mathbf{q})\, \varrho_{m_c\, m_c}(\mathbf{k})  \; ,
\end{eqnarray}
\end{widetext}
with a similar expression for the in-scattering rate $\Gamma^{\text{in}\, \text{X}}_{m_c\,m_c}(\mathbf{k})\,$. These expressions represent all contributions to electron-hole scattering by Coulomb exchange interaction in second order Born approximation. It is important to note that without a macroscopic occupation of hole states (by doping or optical excitation) this scattering rate vanishes: holes are required for the mutual spin flips of the BAP mechanism. 

The EOM for the off-diagonal entry of the density matrix the expressions can be written in the form
\begin{widetext}
\mbox{}\\
\vspace{-0.6cm}
\begin{eqnarray}
\partial_t{\varrho}_{m_c\,-m_c}(\mathbf{k})\pmb{\vert}_{\text{X}} 
&=& \frac{1}{i\, \hbar}\, \Big[\Sigma^{\text{X}}_{m_c\, -m_c}(\mathbf{k})\, \varrho_{m_c\,-m_c}(\mathbf{k}) - \sum_{\bar{\mathbf{k}}\, \mathbf{q}}\sum_{\substack{m_c'\\  \tilde{m}_v\,\tilde{m}_v'}} \bar{\Sigma}^{\text{X}}_{m_c\, -m_c}(\bar{\mathbf{k}}+\mathbf{q})\,\varrho_{m_c\,-m_c}(\bar{\mathbf{k}} + \mathbf{q})\Big]\; .
\label{scaoff} 
\end{eqnarray}
\end{widetext}
As for the electron-phonon scattering \cite{Lechner:2005}, the first self-energy term in Eq. \eqref{scaoff} is proportional to the absolute squared value of the interaction matrix element $\mathcal{V}^{\text{X}}_{m_c\,\tilde{m}_v'\, m_c'\, \tilde{m}_v}(-\mathbf{k}+\mathbf{q},\, \bar{\mathbf{k}},\, \mathbf{q})$ and can be split up into real and imaginary part connected by \emph{Kramers-Kronig} transformation, where the imaginary part
\begin{widetext}
\mbox{}\\
\vspace{-0.6cm}
\begin{eqnarray}
\nonumber
\Im\{\Sigma^{\text{X}}_{m_c\, -m_c}(\mathbf{k})\} &=&\frac{\pi}{\hbar}\, \sum_{\bar{\mathbf{k}}\, \mathbf{q}}\sum_{\substack{m_c'\\  \tilde{m}_v\,\tilde{m}_v'}} |\mathcal{V}^{\text{X}}_{m_c\,\tilde{m}_v'\, m_c'\, \tilde{m}_v}(-\mathbf{k}+\mathbf{q},\, \bar{\mathbf{k}},\, \mathbf{q})|^2\, \delta[\epsilon_{m_c'}(-\bar{\mathbf{k}} + \mathbf{q})-\epsilon_{-m_c}(\mathbf{k}) + \epsilon_{\tilde{m}_v}(-\mathbf{k}+ \mathbf{q}) - \epsilon_{\tilde{m}_v'}(\bar{\mathbf{k}})]\\ \nonumber
&&\mbox{} \hspace{1.8cm} 
\times \Big\{[1- \varrho_{\tilde{m}_v'\,\tilde{m}_v'}(-\bar{\mathbf{k}})]\, [1 - \varrho_{m_c'\, m_c'}(-\bar{\mathbf{k}} + \mathbf{q})]\, \varrho_{\tilde{m}_v\, \tilde{m}_v}(\mathbf{k}-\mathbf{q})\\
&&\mbox{} \hspace{2.1cm} 
+ [1 - \varrho_{\tilde{m}_v\, \tilde{m}_v}(\mathbf{k}-\mathbf{q})]\, \varrho_{\tilde{m}_v'\, \tilde{m}_v'}(-\bar{\mathbf{k}})\, \varrho_{m_c'\, m_c'}(-\bar{\mathbf{k}} + \mathbf{q})  \Big\}
\label{T21}
\end{eqnarray}
\end{widetext}
accounts for dephasing due to scattering, while the real part contributes to the renormalization of the eigenenergies. However, the real and imaginary part of the second term $\bar{\Sigma}^{X}_{m_c\, -m_c}(\bar{\mathbf{k}}+\mathbf{q})$ in Eq. \eqref{scaoff} are not connected by Kramers-Kronig theorem, because they are proportional to a product of two complex valued exchange interaction matrix elements. Nevertheless it is possible to sort out two parts, one proportional to principal values and one proportional to the energy conserving $\delta$-functions. In order to derive the expressions corresponding to the ones given in Ref. \onlinecite{Lechner:2005}, we will present here the part proportional to $\delta$-functions denoted as $\bar{\Gamma}^{\text{X}}_{m_c\, -m_c}(\bar{\mathbf{k}}+ \mathbf{q})$
\begin{widetext}
\mbox{}\\
\vspace{-0.6cm}
\begin{eqnarray}
\nonumber
\bar{\Gamma}^{\text{X}}_{m_c\, -m_c}(\bar{\mathbf{k}}+ \mathbf{q}) &=& \frac{\pi}{\hbar}\, 
\mathcal{V}^{\text{X}}_{m_c\,\tilde{m}_v'\, m_c'\, \tilde{m}_v}(-\mathbf{k}+\mathbf{q},\, \bar{\mathbf{k}},\, \mathbf{q})\,\mathcal{V}_{-m_c\,\tilde{m}_v'\, -m_c'\, \tilde{m}_v}(-\mathbf{k}+\mathbf{q},\, -\bar{\mathbf{k}},\, \mathbf{q})\\ \nonumber
&& \times \delta[\epsilon_{m_c'}(\bar{\mathbf{k}} + \mathbf{q}) - \epsilon_{-m_c}(\mathbf{k}) -
\epsilon_{\tilde{m}_v'}(-\bar{\mathbf{k}}) +
\epsilon_{\tilde{m}_v}(-\mathbf{k} + \mathbf{q})] \\ \nonumber
&& \times\Big\{[1 - \varrho_{\tilde{m}_v\, \tilde{m}_v}(- \mathbf{k} + \mathbf{q})]\, [1-\varrho_{-m_c\, -m_c}(\mathbf{k})]\,\varrho_{\tilde{m}_v'\, \tilde{m}_v'}(- \bar{\mathbf{k}}) \\
&&\mbox{} \hspace{0.2cm} + [1 - \varrho_{\tilde{m}_v'\, \tilde{m}_v'}(- \bar{\mathbf{k}})]\, \varrho_{\tilde{m}_v\, \tilde{m}_v}(-\mathbf{k}+\mathbf{q}) \, \varrho_{-m_c\, -m_c}(\mathbf{k}) \Big\} \;.
\label{T22}
\end{eqnarray}
\end{widetext}
With the expressions given in Eqs. \eqref{T11}, \eqref{scaoff},\eqref{T21} and \eqref{T22} we can define the $T_1$ and $T_2$ times of the Bloch equations as in Ref. \onlinecite{Lechner:2005}, with the $T_1$ time being ruled by the in- and out-scattering terms of Eq. \eqref{T11} while the $T_2$ time is given by the imaginary part of the self-energies of Eq. \eqref{scaoff}. The interpretation of these times with respect to spin-relaxation and -dephasing remains the same as  in Ref. \onlinecite{Lechner:2005} apart from the fact that the spin-relaxation mechanism is different.\\
In conclusion, we have derived microscopic expressions for the scattering rates due to electron-hole exchange interaction in a semiconductor QW in the frame of extended SBE. As it turns out, the expressions for these rates show the same qualitative structure as found for carrier-phonon scattering \cite{Lechner:2005}. The particular scattering mechanism considered here is the one responsible for the BAP mechanism of spin relaxation. Thus, the presented results are a microscopic formulation of the BAP spin relaxation in the frame of the extended semiconductor Bloch equations.\\
We thankfully acknowledge financial support from the DFG via Forschergruppe~370 \emph{Ferromagnet-Halbleiter-Nanostrukturen}.  


\bibliography{./promotion}


\end{document}